\documentclass[fleqn,amsmath,amssymb,aps,pra]{revtex4}

\usepackage{graphicx}
\usepackage{longtable}
\usepackage{dcolumn}
\usepackage{tabularx}
\usepackage{booktabs}

\begin{document}

\title{Laser-stimulated electric quadrupole transitions in the
molecular hydrogen ion H$_2^+$}

\author{V.I.~Korobov}
\affiliation{Bogoliubov Laboratory of Theoretical Physics, Joint Institute for Nuclear Research, 141980, Dubna, Russia}

\author{P.~Danev}%
\affiliation{%
 Institute for Nuclear Research and Nuclear Energy, Bulgarian Academy of Sciences,\\
    blvd. Tsarigradsko ch. 72, Sofia 1142, Bulgaria}

\author{D. Bakalov}%
\email{bakal10@abv.bg}
\affiliation{%
 Institute for Nuclear Research and Nuclear Energy, Bulgarian Academy of Sciences,\\
    blvd. Tsarigradsko ch. 72, Sofia 1142, Bulgaria}

\author{S. Schiller}%
\affiliation{%
 Institut f\"{u}r Experimentalphysik, Heinrich-Heine-Universit\"{a}t D\"{u}sseldorf, 40225 D\"{u}sseldorf, Germany}

\date{\today}

\begin{abstract}
 Molecular hydrogen ions are  of metrological relevance due to the
 possibility of precise theoretical evaluation of their
 spectrum and of external-field-induced shifts.
 We report the results of the calculations of the rate of
 laser-induced electric quadrupole transitions between a large set
 of ro-vibrational states of ${\rm H_2^+}$.
 The hyperfine and Zeeman structure of the
 E2 transition spectrum and the effects of the laser polarization
 are treated in detail. We also present the nuclear spin-electron spin coupling constants,
 computed with a precision 10 times higher than previously.
\end{abstract}

\maketitle

\section{\label{sec:H2p}Introduction}

 Molecular hydrogen ions (MHI) are three-body systems that give the
 possibility of precise theoretical evaluation of their spectrum,
 transitions, and external effect shifts
 \cite{KorobovHFS16},\cite{Bakalov2011}.
 Properly selected transitions exhibit weak sensitivity to external fields.
 This feature makes them excellent candidates for frequency
 standards with potential uncertainties at the $10^{-17}$
 fractional level \cite{BakalovPRL,Karr2016}.
 %
 Current and future results from precision
 spectroscopy of MHI, combined with the theroretical prediction of
 transition frequencies, also allows determining several fundamental constants
 of atomic physics, such as particle mass ratios and the Rydberg
 constant \cite{koel,bies}.

 The spectroscopy of electric quadrupole transitions in homonuclear
 molecules has been the subject of many investigations, recently in
 trapped and sympathetically cooled molecular ions (see, e.g.
 \cite{willi} and references therein.)
 The first theoretical study on the electric quadrupole ro-vibrational
 transitions of H$_2^+$ was published by
 by Bates and Poots in 1953 \cite{BatesAndPoots}
 using the two-centre approximation for the wavefunction.
 Posen et al.  \cite{posen} have computed the spontaneous emission rates for all
 ro-vibrational transitions in H$_2^+$, without inclusion of hyperfine
 structure. More accurate
 calculations of the E2 transition amplitudes in H$_2^+$ were performed by
 Pilon and Baye \cite{Pilon2012} and Karr \cite{Karr2014}.
 In \cite{Karr2014}, however, the hyperfine structure of the E2 transition line
 had not been considered quantitatively except for the particular case
 of stretched states.

 Recently we \cite{BakalovPRL} and Karr \cite{Karr2014}
 (see also Ref.~\cite{Karr2016}) have pointed out that the electric quadrupole
 spectroscopy of H$_2^+$ sympathetically cooled by beryllium ions has
 outstanding potential for achieving ultra-high precision.
 In this context, in \cite{Schiller2017} an approach was proposed
 for quantum state preparation of H$_2^+$, which involves
 laser-driven electric quadrupole transitions.

 The strengths of hyperfine-resolved quadrupole transitions for diatomic
 molecules have recently been discussed by Germann and Willitsch \cite{G&W}.
 Specifically, they considered Hund$'$s case (b), and derived
 for this particular case
 general expressions for the line strength in zero-th order of
 perturbation theory in the spin interactions, without
 taking into account the dependence on laser polarization.

 In this work, we present a complete treatment of the electric
 quadrupole transitions of H$_2^+$, including both the spin (hyperfine)
 structure and the effects of magnetic field and laser
 polarization. In comparison with the preceding results, we have
 considered transitions
 between higher excited states with vibration quantum number up to
 $v = 10$.

 In the following, we derive the explicit expressions for the
 interaction of a monochromatic wave with the H$_2^+$ ion in an
arbitrary quantum state, starting with the basics in Sec.
\ref{2a}.
The hyperfine structure of the levels of H$_2^+$ is introduced in
Sec. \ref{2b}, followed by the computation of the energies of the
spin states. A high accuracy was made possible by an improved
computation of the spin-spin coupling coefficients.
A detailed treatment of the transition strengths of the spin
components of a given ro-vibrational transition is worked out in
Sec. \ref{2c}. We pay particular attention to making our results
easily comparable with previous work.
Sec. \ref{sec3} is devoted to the discussion of some examples that
are believed to be of relevance for near-future precision
spectroscopic studies.

\section{Theory}

\subsection{Interaction with an external electromagnetic field}
\label{2a}

In the center-of-mass frame, the non-relativistic Hamiltonian of
${\rm H_2^+}$ is:
\begin{equation}\label{nonrelhamilt}
H^{\rm NR} =
 \frac{\mathbf{P}_1^2}{2m_p}+\frac{\mathbf{P}_2^2}{2m_p}+\frac{\mathbf{P}_e^2}{2m_e}
 +\frac{e^2}{4\pi\varepsilon_0}\left(
      -\frac{1}{r_1}-\frac{1}{r_2}+\frac{1}{r_{12}}\right),
\end{equation}
 where $\mathbf{R}_{1,2}$, $\mathbf{R}_e$ and $\mathbf{P}_{1,2}$, $\mathbf{P}_{e}$
 are the position and momentum operators of the two protons and the electron, respectively,
 $\mathbf{r}_1=\mathbf{R}_e\!-\!\mathbf{R}_1$,
 $\mathbf{r}_2=\mathbf{R}_e\!-\!\mathbf{R}_2$,
 ${\bf r}_{12}={\bf R}_1-{\bf R}_2$, and $m_p$, $m_e$
 are the masses of the proton and the electron.

The interaction Hamiltonian of a system of particles with an
external electromagnetic field is \cite{CT}:
\begin{equation}\label{Hint}
H_{\rm int} =
   -\sum_{\alpha}\frac{Z_{\alpha}e}{m_{\alpha}}\,\mathbf{P}_{\alpha}\!\cdot\!\mathbf{A}
   (\mathbf{R}_{\alpha},t).
\end{equation}
In Eq.(\ref{Hint}) we have kept only the linear terms  in the
vector potential $\mathbf{A}(\mathbf{R},t)$; $e$ is the magnitude
of the electron charge, the summation runs over all three
constituents of H$^+_2$ $\alpha=p_1,p_2,e^-$, and $Z_{\alpha}$ is
the charge of particle $\alpha$ in units of $e$. For a plane wave
with general polarization the electromagnetic vector potential
is:
\[
 \mathbf{A}(\mathbf{R},t)=\mathbf{A}_0 e^{i(\mathbf{k}\cdot\mathbf{R}-\omega t)}+
 \mathbf{A}^{\ast}_0 e^{-i(\mathbf{k}\cdot\mathbf{R}-\omega t)}
\]
and corresponds to electric field
\begin{equation}
 \mathbf{E}(\mathbf{R},t)=\mathbf{E}_0 e^{i(\mathbf{k}\cdot\mathbf{R}-\omega t)}+
 \mathbf{E}^{\ast}_0 e^{-i(\mathbf{k}\cdot\mathbf{R}-\omega t)},\
 \mathbf{E}_0=i\omega\mathbf{A}_0.
\end{equation}

$\mathbf{A}_0$ is a complex vector satisfying $\mathbf{A}_{0}\cdot
\mathbf{k} = 0$. In the long wavelength approximation, we expand
the exponent $e^{\pm i(\mathbf{k}\cdot\mathbf{R}_{\alpha})}$ in
(\ref{Hint}) and keep only the term responsible for the electric
quadrupole transitions:
\begin{equation}
H^{(2)}_{\rm
int}=-\sum_{\alpha}\frac{Z_{\alpha}e}{m_{\alpha}}\mathbf{P}_{\alpha}\cdot
   (i\mathbf{A}_0 e^{-i\omega t}(\mathbf{k}\cdot\mathbf{R}_{\alpha})
   + c.c.).
\label{H^2}
\end{equation}
 By rearranging the terms we rewrite the above expression as a sum of products of
 symmetric or anti-symmetric tensors. The product of antisymmetric tensor
 gives rise to magnetic dipole transitions and will not be considered here.
 The remaining terms are put in the form \cite{CT}:
\begin{equation}\label{H^E2}
H^{(E2)}_{\rm
int}=\frac{i}{\hbar}\sum\limits_{\alpha}\frac{Z_{\alpha}e}{2\omega}\sum\limits_{ij}
        T^{(2)}_{ij}(t)[R_{\alpha i}R_{\alpha j},H^{\rm NR}].
\end{equation}
 Here $T^{(2)}_{ij}(t)=\frac{1}{2}(k_iE_{j}(0,t)+k_jE_{i}(0,t))$ is the
 symmetric part of the tensor product of the electric field at the center-of-mass
 of the system, ${\bf E}(0,t)$, and of the wave vector, ${\bf k}$.
 We shall also make use of the dimensionless, time-independent and complex tensor
 \begin{equation}
 \widehat{T}_{ij}=(\hat{k}_i\hat{\epsilon}_j+
 \hat{k}_j\hat{\epsilon}_i)/2,
 \label{That}
 \end{equation}
 where $\hat{k}$ and $\hat{\epsilon}$ are unit vectors
 along ${\mathbf k}$ and ${\mathbf E}_0$: ${\mathbf k}=k\,\hat{\mathbf k}=
 (\omega/c)\hat{\mathbf k}$, ${\mathbf E}_0=|{\mathbf E}_0|\,\hat{\mathbf \epsilon}$.
 The relation of $T^{(2)}_{ij}(t)$ to $\widehat{T}_{ij}$ reads:
 \begin{equation}
 T^{(2)}_{ij}(t)=k|{\mathbf E}_0|\left(\widehat{T}_{ij}\,e^{-i\omega t}+
 \widehat{T}^*_{ij}\,e^{i\omega t}\right).
 \label{widehat}
 \end{equation}

\subsection{H$_2^+$ hyperfine structure}
\label{2b}

The calculations in this work are done in the total angular
momentum representation with the following coupling scheme of
angular momentum operators:
\begin{equation}\label{coupling}
\mathbf{I} = \mathbf{I}_1+\mathbf{I}_2, \qquad
\mathbf{F}=\mathbf{I}+\mathbf{s}_e, \qquad
\mathbf{J}=\mathbf{L}+\mathbf{F}.
\end{equation}
$\mathbf{I}_{1,2}$ and $\mathbf{s}_e$ are the spin operators of
the two protons and electron, respectively, $\mathbf{L}$ is the
total orbital momentum, and $\mathbf{J}$ is the total angular
momentum.

 The H$_2^+$ molecular ion has a simple hyperfine structure. As a homonuclear
 molecule with fermionic nuclei,
 its state vectors are antisymmetrical with respect to the exchange of the protons.
 This property gives rise to a restriction on the total nuclear spin, $(-1)^I =
 (-1)^L$,
 (for the ground electronic state $1s\sigma_g$), which therefore becomes
 an exact quantum number \footnote{Note that for homonuclear molecules
 with bosonic nuclei these features are modified; for D$_2^+$, in
 particular, the total nuclear spin $I$ is no longer an exact quantum
 number, and mixing of states with $I=0$ and $I=2$ will take place
 in the expansion, analogous of Eq.~(\ref{realFunction})
 }.
 The other exact quantum numbers are $J$ and the $z$-axis projection $J_z$.
 Although $F$ is not conserved, it can be used as a label of the hyperfine states
 since the mixing in $F$ is small (see the table in the Supplemental material \cite{suppl1}).
 The states with odd orbital quantum number $L$ are split into six
 hyperfine components: $(F,J)=(1/2,L\pm1/2)$, $(3/2,L\pm1/2)$, $(3/2,L\pm3/2)$,
 and for even $L$ --- into two components: $(F,J)=(1/2,L\pm1/2)$.
 Exceptions are the $L = 0$ state with a single component
 $(F,J)=(1/2,L+1/2)$,
 and $L = 1$, which has five components: $(F,J)=(1/2,L\pm1/2)$,
 $(3/2,L\pm1/2)$, $(3/2,L+3/2)$.

 The hyperfine energies $E^{\rm hfs}_{(vL)FJ}$ and state vectors
 are calculated by diagonalization of the effective state-dependent spin
 Hamiltonian $H^{\rm eff}$, obtained from the Breit-Pauli interaction
 by averaging over space variables with the non-relativistic
 wave functions of H$^+_2$ \cite{Korobov2006,Korobov2008},
 \begin{equation}
 \begin{array}{@{}l}\displaystyle
 H^{\rm eff} =
   b_f(\mathbf{I}\cdot \mathbf{s}_e)
   +c_e(\mathbf{L}\cdot \mathbf{s}_e)
   +c_I(\mathbf{L}\cdot \mathbf{I})
   +\frac{d_1}{(2L\!-\!1)(2L\!+\!3)}\left(
      \frac{2}{3}{\bf L}^2({\bf I}\cdot {\bf s}_e)-[({\bf L}\cdot {\bf I})({\bf L}\cdot {\bf s}_e)
      +({\bf L}\cdot {\bf s}_e)({\bf L}\cdot {\bf I})]
   \right)
\\[3mm]\hspace{60mm}\displaystyle
   +\frac{d_2}{(2L\!-\!1)(2L\!+\!3)}\left(
      \frac{1}{3}{\bf L}^2{\bf I}^2-\frac{1}{2}({\bf L}\cdot{\bf I})-({\bf L}\cdot{\bf I})^2
   \right).
\end{array}
\label{H^eff}
\end{equation}
 The state-dependent coefficients $b_f$, $c_e$, $c_I$, $d_1$,
 and $d_2$ are calculated numerically. In this paper we use the recently
 updated values of $b_f$, in which the contributions of order
 $O(m\alpha^6)$, amounting to $10^{-4}$ fractionally,
 have been accounted for. For the remaining
 coefficients we use the values calculated in \cite{Korobov2006}.
 The magnitude of the coefficient $b_f$ dominates over the others,
 which justifies the choice of angular momentum coupling given in Eq.~(\ref{coupling})
 rather than Hund's case (b).
 $H^{\rm eff}$ is an operator acting in the space of spin
 variables and total orbital angular momentum $L$, which is spanned by the basis vectors
\begin{equation}\label{basestates}
|LIFJJ_z\rangle=
   \sum\limits_{\zeta_1\zeta_2 I_z\zeta_eF_zL_z}
   C_{I_1\zeta_1I_2\zeta_2}^{II_z}C_{II_zs_e\zeta_e}^{FF_z}C_{FF_zLL_z}^{JJ_z}
   \times|I_1\zeta_1\rangle|I_2\zeta_2\rangle|{\bf
   s}_e\zeta_e\rangle|LL_z\rangle.
\end{equation}
 Here $|LL_z\rangle$ satisfies
 $(\mathbf{L}^2-L(L+1))|LL_z\rangle=0$, $(\mathbf{L}_z-L_z)|LL_z\rangle=0$,
 and similar for the individual particle spin operators and eigenvectors.
 In first order of perturbation theory the
 hyperfine state vectors, $|(v L)FJJ_z\rangle$,
 are expressed as
 linear combinations of the basis vectors $\bigr|LIFJJ_z\rangle$:
 \begin{align}
 \label{realFunction}
  |(vL)FJJ_z\rangle = \sum\limits_{F'}\beta_{F'}^{(vL)FJ}|LIF'JJ_z\rangle.
 \end{align}
 where $\beta_{F'}^{(vL)FJ}$ are the eigenvectors of
 the matrix of $H^{\rm eff}$ in the basis set (\ref{basestates}),
 and the values of $F'$ satisfy the inequalities
 $\max(|I-1/2|,|J-L|)\le F'\le\min(I+1/2,J+L)$ .
 In absence of external fields, the energy levels are degenerate in $J_z$.
 The non-zero components of $\beta^{(vL)FJ}_{F'}$ can be parameterized as follows:
 for odd $L$ ($I=1$),
 $\beta^{(vL)3/2L\pm3/2}_{3/2}=1,
 \beta^{(vL)FL\pm1/2}_F=\cos\phi_{\pm}, F=1/2,3/2,$
 $\beta^{(vL)1/2L\pm1/2}_{3/2}=
 -\beta^{(vL)3/2L\pm1/2}_{1/2}=\sin\phi_{\pm}$;
 for even values of $L$ ($I=0$), $\beta^{(vL)1/2,L\pm1/2}_{1/2}=1,
 \ \beta^{(vL)1/2,L\pm1/2}_{3/2}=0$.
 The energies $E^{\rm hfs}_{(vL)FJ}$, the mixing angles $\phi_{\pm}$ between spin
 basis states, and the coefficients $b_f$
 are given in Table \ref{tab:suppl1}
 for vibrational and rotational quantum
 numbers in the range $0\le v\le 4$ and $0\le L\le 8$.
 The small values of the mixing angles confirm the appropriateness of
 the coupling scheme of (\ref{coupling}) for the classification of
 the hyperfine structure of the ro-vibrational spectrum of H$_2^+$
 and justify the use -- in lower accuracy estimates -- of the
 zero-th order approximation for the hyperfine state vectors,
 which reads:
 \begin{equation}
 \phi_{\pm}\approx0,\ \beta^{(vL)FJ}_{F'}\approx\delta_{FF'}.
 \label{zero-th}
 \end{equation}

 \begin{center}
  \begin{table}[h]
  \caption{Hyperfine structure of the lower ro-vibrational states
  of H$_2^+$ with orbital
  momentum $L$ in the range $0\le L\le 4$ and vibrational quantum number
  $v$ in the range $0\le v\le 8$.
  Listed are: the updated values (in MHz) of the coefficient
  $b_f$ in the effective spin Hamiltonian
  $H^{\rm eff}$ (\ref{H^eff}), the hyperfine energies
  $E^{\rm hfs}_{(vL)FJ}$ (in MHz) and, for odd $L$, the mixing angles
  $\phi_{\pm}$ (in rad).
  }
  \label{tab:suppl1}
  \begin{footnotesize}
  \begin{tabular}{l c c@{\hspace{4mm}} c c c c c c c c c c c c c c}
  \hline\hline\\
  & & \multicolumn{1}{l}{$b_f$, MHz}
  & \multicolumn{6}{c}{$E^{\rm hfs}_{(vL)FJ}$, MHz}
  &  \multicolumn{2}{c}{mixing angles} & &
  & \multicolumn{2}{c}{$E^{\rm hfs}_{(vL)FJ}$, MHz} \\ \\
  $L$ &$v$& \multicolumn{1}{r}{$(F,J):$}
  & $\left(\frac{3}{2},L+\frac{3}{2}\right)$&$\left(\frac{3}{2},L+\frac{1}{2}\right)$
  &$\left(\frac{1}{2},L+\frac{1}{2}\right)$
  &$\left(\frac{3}{2},L-\frac{1}{2}\right)$&$\left(\frac{1}{2},L-\frac{1}{2}\right)$
  &$\left(\frac{3}{2},L-\frac{3}{2}\right)$&$\phi_{+}$&$\phi_{-}$
  &$L$ &$v$&$\left(\frac{1}{2},L+\frac{1}{2}\right)$&$\left(\frac{1}{2},L-\frac{1}{2}\right)$\\[0.5mm]
 \hline\\
 1&0&922.9318&474.0763&481.9234&$-$930.3732&385.3687&$-$910.6980&&$-$0.01561&$-$0.03890 \vline&2&0&42.1625&$-$63.2438\\
 1&1&898.7507&461.2282&468.4956&$-$905.7253&377.9657&$-$887.1909&&$-$0.01508&$-$0.03736 \vline&2&1&39.5716&$-$59.3574\\
 1&2&876.3973&449.3273&456.0493&$-$882.9269&371.2588&$-$865.4856&&$-$0.01452&$-$0.03578 \vline&2&2&37.0992&$-$55.6487\\
 1&3&855.7571&438.3119&444.5191&$-$861.8606&365.2188&$-$845.4716&&$-$0.01395&$-$0.03417 \vline&2&3&34.7295&$-$52.0943\\
 1&4&836.7296&428.1272&433.8475&$-$842.4234&359.8226&$-$827.0524&&$-$0.01335&$-$0.03253 \vline&2&4&32.4479&$-$48.6718\\
 1&5&819.2274&418.7256&423.9833&$-$824.5255&355.0518&$-$810.1443&&$-$0.01273&$-$0.03084 \vline&2&5&30.2400&$-$45.3600\\
 1&6&803.1750&410.0651&414.8819&$-$808.0888&350.8940&$-$794.6756&&$-$0.01209&$-$0.02911 \vline&4&0&82.5884&$-$103.2355\\
 1&7&788.5079&402.1094&406.5042&$-$793.0463&347.3416&$-$780.5856&&$-$0.01142&$-$0.02733 \vline&4&1&77.4966&$-$96.8707\\
 1&8&775.1714&394.8271&398.8163&$-$779.3412&344.3925&$-$767.8240&&$-$0.01071&$-$0.02548 \vline&4&2&72.6350&$-$90.7937\\
 3&0&917.5313&507.2270&489.4960&$-$941.0438&423.6046&$-$894.6020&341.5241&$-$0.04213&$-$0.06186 \vline&4&3&67.9728&$-$84.9660\\
 3&1&893.6964&492.3526&475.5481&$-$915.6828&413.6522&$-$871.9908&336.8955&$-$0.04067&$-$0.05948 \vline&4&4&63.4807&$-$79.3509\\
 3&2&871.6711&478.5158&462.6063&$-$892.2046&404.5273&$-$851.1450&332.8336&$-$0.03916&$-$0.05705 \vline&&&&\\
 3&3&851.3432&465.6422&450.6022&$-$870.4878&396.1856&$-$831.9595&329.3264&$-$0.03761&$-$0.05456 \vline&&&&\\
 3&4&832.6144&453.6659&439.4752&$-$850.4255&388.5894&$-$814.3434&326.3667&$-$0.03599&$-$0.05200 \vline&&&&\\
 3&5&815.3996&442.5281&429.1720&$-$831.9247&381.7075&$-$798.2185&323.9516&$-$0.03432&$-$0.04937 \vline&&&&\\
 3&6&799.6258&432.1751&419.6456&$-$814.9040&375.5153&$-$783.5195&322.0852&$-$0.03257&$-$0.04664 \vline&&&&\\
 3&7&785.2282&422.5607&410.8538&$-$799.2920&369.9910&$-$770.1888&320.7713&$-$0.03074&$-$0.04382 \vline&&&&\\
 3&8&772.1561&413.6433&402.7612&$-$785.0303&365.1210&$-$758.1833&320.0235&$-$0.02882&$-$0.04090 \vline&&&&\\
 \hline\hline
 \end{tabular}
 \end{footnotesize}
 \end{table}
 \end{center}

 \subsection{E2 transition matrix elements and transition rates}
 \label{2c}

 Using Eqs.~(\ref{H^E2}) and (\ref{widehat}), 
 the E2 transition matrix element between
 initial $|i\rangle = \ \bigr|(vL)FJJ_z\rangle$ and final
 $|f\rangle = \ \bigr|(v'L')F'J'J_z'\rangle$
 hyperfine states of H$_2^+$ can be put in a form that
 exhibits the dependence on time:
  \begin{eqnarray}
 &&\langle (v'L')F'J'J_z'\bigl| H^{(E2)}_{\rm int}\bigr|(vL)FJJ_z\rangle
 %
 =\frac{i}{3}\frac{\omega^{\rm NR}}{\omega}\,
 \langle (v'L')F'J'J_z'| T^{(2)}(t)\cdot Q^{(2)}|(vL)FJJ_z\rangle
 %
 \nonumber\\
 &&=\frac{i}{3c}\omega^{\rm NR}\,|{\mathbf E}_0|
 \left(
 e^{-i\omega t}
 \langle (v'L')F'J'J_z'| \widehat{T}\cdot
 Q^{(2)}|(vL)FJJ_z\rangle+
 e^{i\omega t}
 \langle (v'L')F'J'J_z'| \widehat{T}^*\!\cdot
 Q^{(2)}|(vL)FJJ_z\rangle
 \right).
 \label{H^E2mel1}
 \end{eqnarray}
 Here $\omega^{\rm NR} = (E^{\rm NR}_{v'L'}-E^{\rm NR}_{vL})/\hbar$,
 $E^{\rm NR}_{vL}$ and $E^{\rm NR}_{v'L'}$ are the
 non-relativistic energies of the
 initial and final states, $Q^{(2)}$ is the irreducible tensor of the
 electric quadrupole moment of H$_2^+$
 \begin{align}
  &Q_{ij} = \frac{3}{2}\sum\limits_{\alpha}Z_{\alpha}e
  \left( R_{i \alpha}R_{j \alpha} - \frac{1}{3} ({\bf R}_{\alpha})^2\delta_{ij}
  \right),
 \end{align}
 and $T^{(2)}(t)\cdot Q^{(2)}\equiv\sum\limits_{ij} T^{(2)}_{ij}(t)
 Q^{(2)}_{ij}$ denotes the scalar product of the tensors.
 The cyclic components $Q^{(2)}_q$ and $T^{(2)q}(t)$, $q=-2,\ldots,2$
 are normalized by $Q^{(2)}_0 = Q_{zz}$ and similar for $T^{(2)0}$
 (cf. \cite{Karr2014,James}).
 In terms of the cyclic components the scalar product is expresses
 as $T\cdot Q^{(2)} =
 \frac{3}{2}\sum\limits_{q} T^{(2)q}(t)Q_{q}^{(2)}$.
 Some authors use alternative normalization conventions (e.g.
 \cite{Bakalov2014} and \cite{wars}); the current convention was selected to ease
 comparison with the numerical results of \cite{Pilon2012} and
 \cite{Karr2014}.
 In these notations the Rabi frequency of the
 $|(vL)FJJ_z\rangle\rightarrow|(v'L')F'J'J'_z\rangle$
 transition is given by
 \cite{James}
 \begin{equation}
 \Omega_{if}=
 \frac{|{\mathbf E}_0|\,\omega_{if}}{3\hbar c}
 \left|\langle
 (v'L')F'J'J_z'|\,\widehat{T}\!\cdot\!Q^{(2)}\,|(vL)FJJ_z\rangle
 \right|.
 \label{Rabi}
 \end{equation}
 Next, using Eq.~(\ref{realFunction}), for the time-independent matrix
 elements of $\widehat{T}\cdot Q^{(2)}$ we have
 \begin{equation}
 \langle (v'L')F'J'J_z'| \widehat{T}\cdot
 Q^{(2)}|(vL)FJJ_z\rangle=\frac{3}{2}\sum\limits_q
 \widehat{T}^q \sum\limits_{F_1,F'_1}
 \beta^{(vL)FJ}_{F_1} \beta^{(v'L')F'J'}_{F'_1}
 \langle
 L'IF'J'J'_z\bigl|Q_{q}^{(2)}|LIFJJ_z\rangle.
 \end{equation}
 With the Wigner-Eckart theorem the matrix elements of
 $Q^{(2)}$ in the basis set (\ref{basestates}) are
 expressed in terms of the non-relativistic reduced matrix
 elements $\langle v'L'\|Q^{(2)}\|vL\rangle$:
 \begin{align}
 &\langle
 L'IF'J'J'_z\bigl|Q_{q}^{(2)}|LIFJJ_z\rangle=\delta_{F'F}
 (-1)^{J+L+F}\sqrt{2J+1}\,C_{JJ_z,2q}^{J'J'_z}
 \left\{\begin{matrix}
      L & \!F\! & J \\
      J' & \!2\! & L'
   \end{matrix}\right\}
   \langle v'L'\|Q^{(2)}\|vL\rangle.
 \nonumber
 \end{align}
 The above expression implies the selection rules $\Delta L
 \equiv L'-L=0,\pm2$, $|\Delta J|\le2$, $J+J'\ge2$; the transitions $L=0\rightarrow
 L'=0$ are also forbidden.

 The probability ${\mathcal W}_{if}(T)$ for a particular E2 transition
 $|i\rangle\equiv|(vL)IFJJ_z\rangle\rightarrow|f\rangle\equiv|(v'L')IF'J'J'_z\rangle$
 in the time interval $0\le t\le T, \omega T\gg1$,
 stimulated by the external electromagnetic field ${\mathbf E}$
 oscillating with frequency $\omega$, is \cite{CT}:
 \begin{equation}
 {\mathcal W}_{if}(T)=
 \left|\frac{1}{\hbar}\int\limits_0^T dt\,\exp (i\omega_{if}t)\,
 \langle (v'L')F'J'J_z'\bigl| H^{(E2)}_{\rm int}\bigr|(vL)FJJ_z\rangle
 \right|^2\approx\Omega^2_{if}
 \left(F((\omega\!-\!\omega_{if});T)+F((\omega\!+\!\omega_{if});T)\right),
 %
 \label{W_{if}(Tcap)}
 \end{equation}
 where $\omega_{if}=(E_f-E_i)/\hbar$ is the transition angular frequency with
 account of the hyperfine, Zeeman etc. corrections to $\omega^{\rm
 NR}$, and the following $\delta$-like function is used:
 $F(a;T)=(\sin(aT/2)/(a/2))^2$, $\lim\limits_{T\to\infty}=2\pi T\delta(a)$.
 The rate ${\mathcal W}_{if}$ of the transition is defined as the probability
 per unit time over a sufficiently long time interval:
 ${\mathcal W}_{if}={\mathcal W}_{if}(T)/T,\ \omega\,T\gg1$.
 We shall put the expression of ${\mathcal W}_{if}$ in a form
 that accounts for the characteristics of the laser source and
 the transition line  profile, and distinctly exhibits the
 hyperfine and Zeeman structure of the spectrum.
 To this end we relate the amplitude ${\mathbf E}_0$ of the
 electric field to the spectral density of the laser energy flux
 ${\mathcal I}(\omega)$: ${\mathcal I}(\omega)=(\varepsilon_0c/2)\,(d|{\mathbf
 E}_0|^2/d\omega)$, normalized to the laser intensity ${\mathcal I}_0$
 by $\int d\omega\,{\mathcal I}(\omega)= {\mathcal I}_0$.
 We also denote by $g_{if}(\omega)$ the transition line spectral profile
 (determined by Doppler broadening or else), with normalization
 $\int d\omega\,g_{if}(\omega)=1$.
 By combining Eqs.~(\ref{H^E2mel1})-(\ref{W_{if}(Tcap)}),
 in the limits of large $T$,
 the expression for ${\mathcal W}_{if}$ is cast in the
 following factorized form:
 \begin{equation}
 {\mathcal W}_{if}={\mathcal W}^{\rm NR}(v'L';vL)\,
 {\mathcal W}^{\rm hfs}((v'L')F'J';(vL)FJ)\,
 {\mathcal W}^{\rm pol}(J'_z;J_z).\label{fact}
 \end{equation}
 The first factor, ${\mathcal W}^{\rm NR}(v'L';vL)$, is the rate of stimulated
 E2 transitions in H$^+_2$ in the non-relativistic (spinless)
 approximation, averaged over the initial and summed over the final
 angular momentum projections $J_z,J'_z$
 \begin{equation}
 {\mathcal W}^{\rm NR}(v'L';vL)= 
 \frac{\pi\omega_{if}^2}{\varepsilon_0c^3\hbar^2}\frac{1}{15(2L+1)}\left|\langle
 v'L'||Q^{(2)}||vL\rangle\right|^2\, \bar{\mathcal I},
 \ \ \bar{\mathcal I}=\int d\omega\,{\mathcal
 I}(\omega)g_{if}(\omega).
 \label{W^nr}
 \end{equation}
 %
 The factor ${\mathcal W}^{\rm hfs}((v'L')F'J';(vL)FJ)$ is the
 relative intensity of the individual hyperfine components
 $FJ\rightarrow F'J'$ of the transition line
 $(vL)\rightarrow(v'L')$. For simplicity of the notations
 we shall omit $v,L,v',L'$
 wherever possible:
 \begin{eqnarray}
 &&
 {\mathcal W}^{\rm hfs}((v'L')F'J';(vL)FJ)
 \equiv{\mathcal W}^{\rm hfs}(F'J';FJ)
 \label{W^hfs}
 \\
 &&=(2L+1)(2J+1)(2J'+1)
 \left(\sum_{F_1}\beta^{(v'L')F'J'}_{F_1}\beta^{(vL)FJ}_{F_1}
 (-1)^{J+F_1}
  \left\{\begin{matrix}
      L & \!F_1\! & J \\
      J' & \!2\! & L'
   \end{matrix}\right\}
 \right)^2,
 \nonumber
 \end{eqnarray}
 normalized by the condition
 \begin{equation}
 \sum_{F'J'}\frac{1}{n^{\rm hfs}(vL)}\sum_{FJ}
  {\mathcal W}^{\rm hfs}(F'J';FJ)=1,
 \end{equation}
 $n^{\rm hfs}(vL)=2(2I+1)(2L+1)$ standing for the number of
 states $|(vL)FJJ_z\rangle$ of the hyperfine structure of the initial
 ro-vibrational $(vL)$ state. Note that the approximate expression
 for ${\mathcal W}^{\rm hfs}(F'J';FJ)$ that stems from the
 approximation for the amplitudes
 $\beta^{(vL)FJ}_{F'}$ of zero-th order of
 perturbation theory in Eq.~(\ref{zero-th}) 
 \begin{equation}
 {\mathcal W}^{\rm hfs}(F'J';FJ)\approx
 \delta_{FF'}(2L+1)(2J+1)(2J'+1)
  \left\{\begin{matrix}
      L & \!F\! & J \\
      J' & \!2\! & L'
   \end{matrix}\right\}^2
 \label{special_case}
 \end{equation}
 does not describe the ``weak'' hyperfine components of the
 transition lines.

 Finally, ${\mathcal W}^{\rm pol}(J'_z,J_z)$ is the relative intensity of
 the components of the transition line with different values of
 the quantum numbers $J_z, J'_z$ (the dependence on $J$ and $J'$ being omitted for
 simplicity of the notations):
 \begin{align}
 {\mathcal W}^{\rm pol}(J'_z;J_z)=\frac{15}{2J'+1}
 \left(C_{JJ_z,2q}^{J'J'_z}\right)^2
 \left|\widehat{T}^{(2)q}\right|^2,\ q=J'_z-J_z,
 \label{W^pol}
 \end{align}
 expressed in terms of Clebsch-Gordan coefficients and the tensor $\widehat{T}$
 defined in Eq.~(\ref{That}), and satisfying
 \begin{equation}
 \sum_{J_z;J'_z}{\mathcal W}^{\rm pol}(J'_z,J_z)=1,
 \label{sW^pol}
 \end{equation}
 To avoid any ambiguity, we
 list the general expressions of $\widehat{T}^{(2)q}$ in terms of
 the Cartesian components of $\hat{\mathbf k}$ and $\hat{\mathbf \epsilon}$:
 \begin{eqnarray*}
 &&\widehat{T}^{(2)\pm2}=\sqrt{\frac{3}{8}}\left(
 \hat{k}_x\hat{\epsilon}_x-\hat{k}_y\hat{\epsilon}_y
 \mp i\left(\hat{k}_x\hat{\epsilon}_y+\hat{k}_y\hat{\epsilon}_x\right)
 \right)\\
 &&\widehat{T}^{(2)\pm1}=\sqrt{\frac{3}{8}}\left(\mp\left(
 \hat{k}_x\hat{\epsilon}_z+\hat{k}_z\hat{\epsilon}_x\right)+
 i\left(\hat{k}_z\hat{\epsilon}_y+\hat{k}_y\hat{\epsilon}_z\right)
 \right)\\
 &&\widehat{T}^{(2)0}=\frac{1}{2}\left(
 2\hat{k}_z\hat{\epsilon}_z-\hat{k}_x\hat{\epsilon}_x
 -\hat{k}_y\hat{\epsilon}_y\right).
 \end{eqnarray*}

 The relevance of each of the three factors in Eq.~(\ref{fact}) is
 discussed below.

\section{Numerical results and discussion}
\label{sec3}

\subsection{E2 transition rates in the approximation of spin-less particles}
\begin{table}[h]
 \caption{Numerical results for selected E2 transitions in
 H$_2^+$ in the approximation of spin-less particles.
 Comparison with the values of
 Einstein coefficients $A$, calculated by
 Pilon and Baye \cite{Pilon2012}, and of the reduced matrix elements
 $\langle v'L'||Q^{(2)}||vL\rangle$
 of Karr \cite{Karr2014} is made
 in a few illustrative cases considered by those authors.
 The notation $a[b]=a\times10^b$ has been used.}
 \label{tab:comparison_new}
\begin{tabular}{l@{\hspace{3mm}}d@{\hspace{2mm}}
 d@{\hspace{2mm}}d@{\hspace{2mm}}d@{\hspace{2mm}}d}
 \hline\hline \vrule width 0pt height 13pt depth 5pt
 \phantom{$|v$}$i$\phantom{$L\rightarrow|v'$}$f$ & &
 \multicolumn{2}{c}{$\left|\langle\ v'L'\bigr\|Q^{(2)}\bigr\|vL\rangle\right|/ea_0^2$}
 & \multicolumn{2}{c}{Einstein coefficient $A_{fi}$, s$^{-1}$}
 \\
\cline{3-6}
  \multicolumn{1}{l}
  {\vrule width 0pt height 12pt depth 6pt $|vL\rangle\rightarrow|v',L'\rangle$}
  & \multicolumn{1}{c}{\hspace*{-5mm}$\Delta E^{\rm NR}$, cm$^{-1}$}
  & \multicolumn{1}{c}{\hspace*{5mm}This work}
  & \multicolumn{1}{c}{\hspace*{5mm}Karr \cite{Karr2014}}
  & \multicolumn{1}{c}{\hspace*{9mm}This work}
  & \multicolumn{1}{c}{\hspace*{13mm}Pilon\&Baye \cite{Pilon2012}}
  \\
\hline
\vrule width 0pt height 10pt 
 $|0,0\rangle\rightarrow|0,2\rangle$ &174.230 & 1.644960 & &\ 0.973137[-11] & 0.973137[-11]\\
 $|0,0\rangle\rightarrow|1,2\rangle$ & 2356.155 & 0.313846 & &\ 0.160207[-6] & 0.160207[-6]\\
 $|0,0\rangle\rightarrow|4,2\rangle$ & 8156.599 & 0.001048 &  &\ 0.888671[-9] & \\
 $|0,0\rangle\rightarrow|6,2\rangle$ & 11436.092 & 0.000100 &  &\ 0.441771[-10] & \\
 $|0,1\rangle\rightarrow|1,1\rangle$ & 2188.035&\ 0.376163 &\ 0.3762
  & 0.264911[-6]& 0.264911[-6]\\
 $|0,1\rangle\rightarrow|2,1\rangle$ & 4248.965&\ 0.028875 &\ 0.02887
  &\ 0.431045[-7] & 0.431045[-7]\\
 $|0,1\rangle\rightarrow|3,1\rangle$ & 6186.996 &\ 0.004044 &\ 0.004044
  &\ 0.553479[-8] & 0.553479[-8]\\
 $|0,1\rangle\rightarrow|4,1\rangle$ & 8005.682 &\ 0.000792 & & 0.770296[-9] &\\
 $|0,2\rangle\rightarrow|1,2\rangle$ & 2181.925 &\ 0.411812 &\ 0.4118
  &\ 0.185876[-6] & 0.185876[-6]\\
 $|0,2\rangle\rightarrow|2,2\rangle$ & 4236.951 &\ 0.031686 &\ 0.03169
  &\ 0.307072[-7] & 0.307072[-7]\\
 $|0,2\rangle\rightarrow|4,2\rangle$ & 7982.369 &\ 0.000874 & & 0.554230[-9] &\\
 $|0,3\rangle\rightarrow|1,1\rangle$ & 1899.184&\ 0.529496 &\ 0.5295
  &\ 0.258605[-6] & 0.258605[-6]\\
 $|0,4\rangle\rightarrow|1,2\rangle$ & 1780.716 &\ 0.667785 &\ 0.6678
  &\ 0.178844[-6] & 0.178844[-6]\\
 $|0,4\rangle\rightarrow|2,4\rangle$ & 4195.295 &\ 0.041347 &
  &\ 0.276482[-7] & 0.276482[-7]\\
 $|0,4\rangle\rightarrow|3,2\rangle$ & 5768.052&\ 0.002013 &\ 0.002013
  &\ 0.579405[-9]& 0.579407[-9]\\
 $|0,6\rangle\rightarrow|5,6\rangle$ & 9425.184 & 0.000367 &  & 0.864640[-10] & \\
 $|1,1\rangle\rightarrow|1,3\rangle$ & 273.621 & 2.529827 &
  & \ 0.157047[-9] &  0.157047[-9]\\
 $|1,1\rangle\rightarrow|3,3\rangle$ & 4243.606 & 0.074474 &
  &\ 0.122120[-6] & 0.122120[-6] \\
 $|2,0\rangle\rightarrow|4,2\rangle$ & 3901.609 & 0.080164 &  & \ 0.130139[-6] & \\
 $|3,2\rangle\rightarrow|6,4\rangle$ & 5375.331 & 0.049248 &  & \ 0.135446[-6] & \\
 $|5,2\rangle\rightarrow|5,4\rangle$ & 301.389 & 4.850333 &  & \ 0.728017[-9] & \\
 $|4,3\rangle\rightarrow|10,5\rangle$ & 8708.227 & 0.006665 &  & \ 0.226499[-7] & \\
 $|9,6\rangle\rightarrow|10,6\rangle$ & 1087.942 & 3.085308 &  & \ 0.124992[-6] & \\
 \hline\hline
 \end{tabular}
 \end{table}

 The computational challenge in the present paper was the evaluation
 of the reduced matrix elements of the electric quadrupole moment
 of H$_2^+$ in Eq.~(\ref{W^nr}).
 The values of $\langle v'L'||Q^{(2)}||vL\rangle$ were calculated with
 the variational wave functions obtained in the approach of
 \cite{KorobovVarMethod}.

 Some authors present the rates of
 transitions in terms of the Einstein coefficients $A$
 rather than ${\mathcal W}_{if}$,
 related to the reduced matrix elements by \cite{Pilon2012}
 \begin{equation}
 \left(A_{fi}/t_0^{-1}\right)=\frac{\alpha^5}{15(2L+1)}
 \left((E^{\rm NR}_{v'L'}-E^{\rm NR}_{vL})/{\mathcal E}_0\right)^5
 \left(\langle v'L'||Q^{(2)}||vL\rangle/(ea_0^2)\right)^2,
 \label{relat}
 \end{equation}
 where $a_0$, $t_0=a_0/\alpha c$, and ${\mathcal E}_0=2{\rm Ry}$ are the
 atomic units of length (i.e. the Bohr radius), time and energy.
 The numerical results, including the non-relativistic E2 transition energies,
 the values of the reduced matrix elements and of Einstein's
 coefficients
 for {\em all} E2 transitions between states
 with vibrational quantum number $0\le v\le10$, $|v'-v|\le6$, and total orbital
 momentum $0\le L \le6$
 are given in the Table of the
 supplemental material \cite{suppl2}.
 The considered transitions belong to the near
 and mid infrared spectral range; they are to some extent
 complementary to the set of states considered in previous
 works \cite{Pilon2012,Karr2014} and include, among other,
 higher vibration excitations.
 The results are intended to help
 select transitions of appropriate wave length and intensity and
 plan the future experiments on E2 spectroscopy of H$_2^+$.

 Table \ref{tab:comparison_new} illustrates on a few examples
 the agreement between the numerical values of the reduced matrix elements of the
 ${\rm H_2^+}$ electric quadrupole moment
 calculated in \cite{Pilon2012,Karr2014} and in the present work.
 In the overlapping cases the
 numerical results agree within the claimed precision of six
 digits with exceptions that should be attributed
 to the different ways of rounding.

\subsection{Hyperfine structure of the E2 transition spectrum}

\begin{figure}[h]
 \begin{center}
 \includegraphics[width=0.7\textwidth]{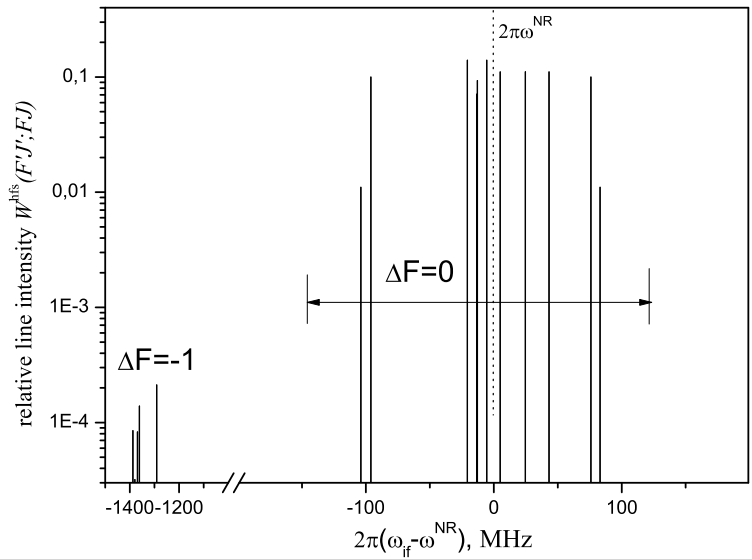}
 \caption{Hyperfine structure of the $(0,1)\rightarrow(1,1)$ E2
 transition line: relative intensity ${\mathcal W}^{\rm
 hfs}(F'J';FJ)$ of the individual hyperfine
 components $(FJ)\rightarrow(F'J')$,
 calculated using Eq.~(\ref{W^hfs}). The ``weak'' components
 with $\Delta F=-1$ around $2\pi(\omega_{if}-\omega^{\rm NR})\sim -1300$ MHz,
 as well as those with $\Delta F=1$ around 1300 MHz (not shown on the plot)
 are suppressed by approximately 3 orders of
 magnitude. The ``strong'' components with $\Delta F=0$ are spread
 over an interval of 200 MHz width around the center of the
 hyperfine structure manifold.}\label{fig:hfs1}
 \end{center}
 \end{figure}

 The non-relativistic picture, in which the E2 transition spectral
 lines are labeled with the quantum numbers $(v,L)$,
 $(v',L')$ of the initial and final states of H$_2^+$, is
 applicable only if the spectroscopic resolution is $\simeq 1\,$GHz
 or worse.
 Under higher resolution, the line will evidence splitting into
 a set of hyperfine components
 spread over an interval of the order of 1 GHz around the
 non-relativistic transition frequency $\omega^{\rm NR}$
 (see Fig.~\ref{fig:hfs1}).
 Transitions having $\Delta F\equiv F'-F=\pm1$ are strongly suppressed compared
 with transitions with $\Delta F=0$;
 the latter are spread over a much narrower
 frequency interval of the order of 100 MHz.
 In the assumption of a flat laser spectral profile ${\mathcal
 I}(\omega)$
 the sum of the transition rates (also referred to as line intensities)
 of all the hyperfine lines equals the non-relativistic intensity
 ${\mathcal W}^{\rm NR}(v'L';vL)$, Eq.~(\ref{W^nr}).
 The relative intensity of the hyperfine components
 is given by the factor
 ${\mathcal W}^{\rm hfs}_{if}$ in Eqs.~(\ref{fact},\ref{W^hfs}).

 Table \ref{tab:intesities} lists the
 energy shift $\Delta E^{\rm hfs}$ and relative intensity
 ${\mathcal W}^{\rm hfs}(F'J';FJ)$ of the
 strong components of a few E2
 transitions of potential interest for precision
 spectroscopy. Note the much simpler structure of the
 ro-vibrational transitions between levels with even $L$.
\begin{table}[ht]
\caption{Hyperfine shifts $\Delta E^{\rm hfs}= E^{\rm
hfs}((v'L')FJ')-E^{\rm hfs}((vL)FJ)$, in MHz, and  relative
intensities ${\mathcal W}^{\rm hfs}(FJ';FJ)$ of the "strong"
components $(F'=F)$ in the hyperfine structure of selected E2
transitions in ${\rm H_2^+}$. The transitions marked in bold have
been identified in Refs. \cite{BakalovPRL,Karr2016}
 as being of particular interest in precision
spectroscopy.}
 \label{tab:intesities}
\begin{center}
\begin{tabular}{c@{\hspace{12mm}}c@{\hspace{12mm}}c@{\hspace{12mm}}r@{\hspace{12mm}}c}
\hline\hline
 $F$ & $J$ & $J'$ &
    \multicolumn{1}{c}{\hspace*{-10mm}$\Delta E^{\rm hfs}$, MHz} &
    \multicolumn{1}{c}{${\mathcal W}^{\rm hfs}(F'J';FJ)$}\\
\hline
   \multicolumn{5}{c}{\vrule width 0pt height 10.5pt depth 5.5pt $(v L) = (00)
   \rightarrow (v' L') = (12)$}\\
    \hline
  {\bf 1/2} & {\bf 1/2} & {\bf 3/2} & \textbf{-59.35740} & \textbf{0.400000} \\
  {\bf 1/2} & {\bf 1/2} & {\bf 5/2} & \textbf{39.57160} &  \textbf{0.600000} \\
\hline
     \multicolumn{5}{c}{\vrule width 0pt height 10.5pt depth 5.5pt $(v L) = (01)
     \rightarrow (v' L') = (11)$}\\
 \hline
 3/2 & 3/2 & 1/2 & \textrm{-103.95790} & \textrm{0.011052}  \\
 3/2 & 5/2 & 1/2 & \textrm{ -96.11068} & \textrm{0.099861}  \\
  3/2 & 3/2 & 5/2 & \textrm{-20.69538} & 0.139966\\
  3/2 & 3/2 & 3/2 & \textrm{-13.42782} & 0.071036\\
 {\bf 3/2} & {\bf 5/2} & {\bf 5/2} & \textbf{-12.84816} & \textbf{0.093333}\\
  3/2 & 1/2 & 1/2 & \textrm{-7.40299} & 0.000000\\
  3/2 & 5/2 & 3/2 & \textrm{-5.58060} & 0.139968\\
  1/2 & 1/2 & 3/2 & \textrm{4.97277} & 0.110877\\
  1/2 & 1/2 & 1/2 & \textrm{23.50717} & 0.000000\\
  {\bf 1/2} & {\bf 3/2} & {\bf 3/2} & \textbf{24.64797} & \textbf{0.111017}\\
  1/2 & 3/2 & 1/2 & \textrm{43.18237} & 0.110888\\
  3/2 & 1/2 & 5/2 & \textrm{75.85953} & 0.099849\\
  3/2 & 1/2 & 3/2 & \textrm{83.12709} & 0.011051\\
\hline
   \multicolumn{5}{c}{\vrule width 0pt height 10.5pt depth 5.5pt $(v L) = (02)
   \rightarrow (v' L') = (12)$}\\
\hline
    1/2 & 5/2 & 3/2 & -101.51990 & 0.120000\\
    {\bf 1/2} & {\bf 5/2} & {\bf 5/2} & \textbf{-2.59090} & \textbf{0.480000}\\
    {\bf 1/2} & {\bf 3/2} & {\bf 3/2} & \textbf{3.88635} & \textbf{0.280000}\\
    1/2 & 3/2 & 5/2 & 102.81535 & 0.120000\\
\hline
   \multicolumn{5}{c}{\vrule width 0pt height 10.5pt depth 5.5pt $(v L) = (03)
   \rightarrow (v' L') = (13)$}\\
\hline
    3/2 & 7/2 & 3/2 & -152.60051 & 0.005433\\
    3/2 & 9/2 & 5/2 & -93.57469  & 0.004237\\
    3/2 & 5/2 & 3/2 & -86.70920  & 0.043913\\
    3/2 & 7/2 & 5/2 & -75.84373  & 0.057349\\
    3/2 & 9/2 & 7/2 & -31.67871  & 0.051879\\
    1/2 & 5/2 & 7/2 & -21.08081  & 0.020469\\
    3/2 & 9/2 & 9/2 & -14.87429  & 0.181878\\
    3/2 & 7/2 & 7/2 & -13.94774  & 0.075715\\
    3/2 & 5/2 & 5/2 & -9.95241   & 0.037264\\
    3/2 & 3/2 & 3/2 & -4.62881   & 0.045714\\
    3/2 & 7/2 & 9/2 & 2.85668    & 0.051873\\
    1/2 & 5/2 & 5/2 & 22.61118   & 0.122044\\
    1/2 & 7/2 & 7/2 & 25.36093   & 0.169874\\
    3/2 & 5/2 & 7/2 & 51.94357   & 0.057341\\
    3/2 & 5/2 & 9/2 & 68.74799   & 0.004235\\
    1/2 & 7/2 & 5/2 & 69.05292   & 0.020471\\
    3/2 & 3/2 & 5/2 & 72.12798   & 0.043926\\
    3/2 & 3/2 & 7/2 & 134.02396  & 0.005433\\
 \hline\hline
\end{tabular}
\end{center}
\end{table}

 \subsection{Laser polarization effects}

\begin{figure}[t]
\begin{center}
     \includegraphics[width=0.8\textwidth]{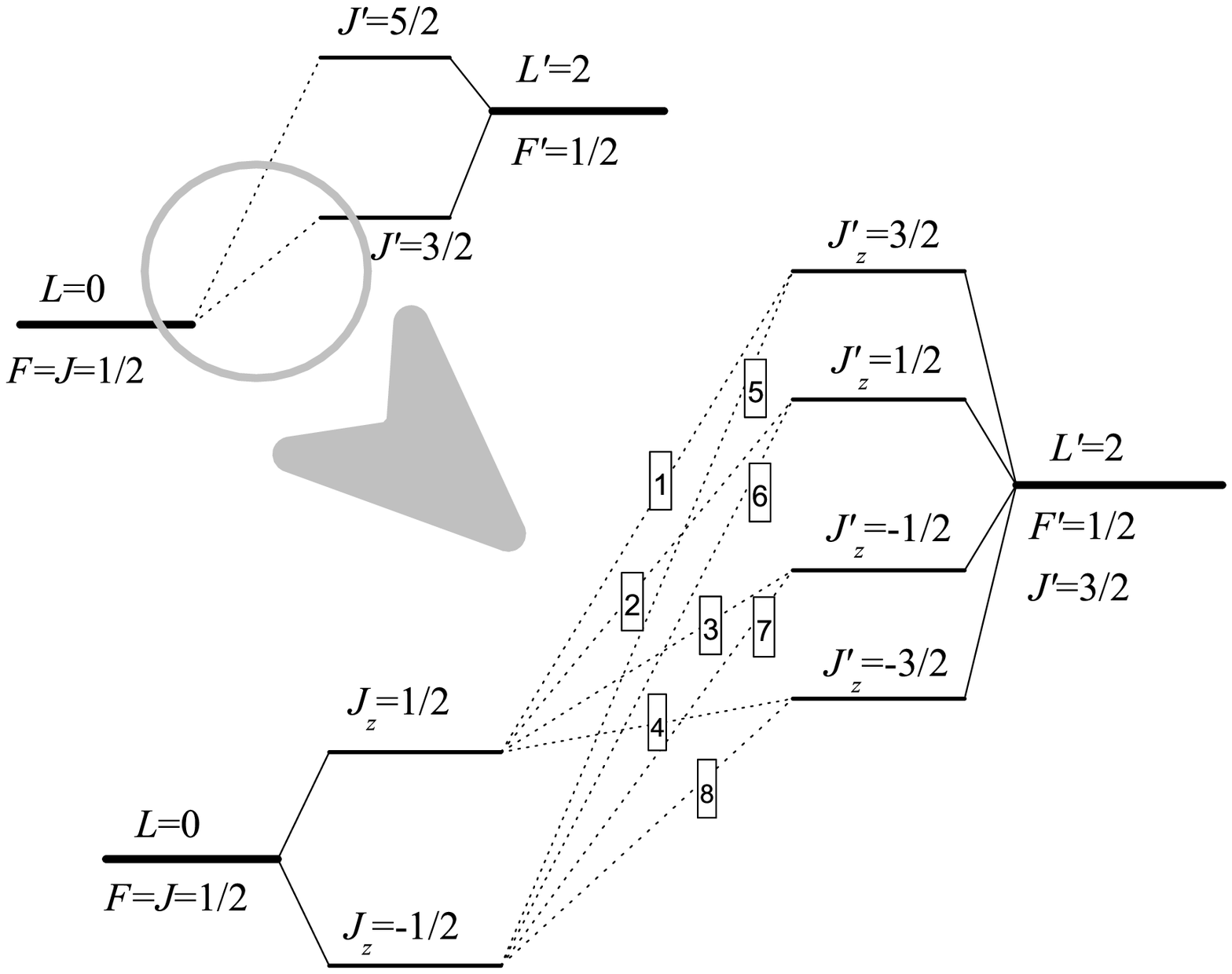}
      \caption{ Zeeman structure of a
      $(v,L=0)\rightarrow(v',L=2)$ transition line of ${\rm H_2^+}$.
      In presence of external magnetic field the
      $(F=J=1/2)\rightarrow(F'=1/2,J'=3/2)$ hyperfine transition line
      (circled in the upper panel) is split into
      8 Zeeman components (lower panel) whose intensities,
      according to Eqs.(\ref{fact}),(\ref{W^pol}), depend
      strongly on geometry and on the polarization of the laser
      radiation. For details, see also Table \ref{tab:zee}.}\label{fig:zee}
\end{center}
\end{figure}

 If the Zeeman structure is not resolved, for example if the magnetic
 field strength is small,  Doppler broadening is large, and the $J_z$-states
 are equally populated, the spectrum is independent of the polarization
 state of the driving laser field, as expressed by Eq.~\ref{sW^pol}.
 Very-high (sub-MHz) resolution spectroscopy of hydrogen molecular
ions could further distinguish the individual Zeeman-split
components of the E2 transition lines. In that case, the
polarization state of the driving laser field becomes relevant,
through the factor ${\mathcal W}^{\rm pol}$. Each hyperfine
transition line $((vL)FJ)\rightarrow((v'L')F'J')$ is split into a
large number of components
$((vL)FJJ_z)\rightarrow((v'L')F'J'J'_z)$.
The Zeeman structure of one E2 transition is illustrated in
Fig.~\ref{fig:zee}. The small number of hyperfine and Zeeman
components of E2 transitions from or to $L=0$ states makes them
particularly appropriate for precision spectroscopy. The linear
and quadratic Zeeman shifts have been calculated precisely and
shown to be of the order of 1 kHz in a field of 1 Gauss for
selected transitions \cite{BakalovPRL,zee-karr}.

 \begin{table}[t]
 \caption{Relative intensities ${\mathcal W}^{\rm pol}(J_z,J'_z)$ of the
 Zeeman components of the $(v,L=0,F=J=1/2)\rightarrow (v',L'=2,F'=1/2,J'=3/2)$
 hyperfine transition, for linear and
 circular polarization of the incident light and a set of angles $\beta$ between
 the directions of the laser beam and the magnetic field
 (see Fig.~\ref{fig:zee}).
 The values of ${\mathcal W}^{\rm pol}(J_z,J'_z)$ were calculated
 from Eq.~(\ref{W^pol}), using expressions
 (\ref{linpol},\ref{circpol}) for the explicit dependence on
 the angle $\beta$.
 Note that the relative intensities depend only on the quantum numbers
 $J,J_z,J'$ and $J'_z$, and are independent of $v,v',L,L'$, the
 transition energy or the magnitude of the Zeeman shift.}
 \label{tab:zee}
 \begin{tabular}{ccc|@{\hspace{6mm}}
 l@{\hspace{3mm}}l@{\hspace{3mm}}l
 @{\hspace{3mm}}l@{\hspace{3mm}}
 |@{\hspace{3mm}}l@{\hspace{3mm}}l@{\hspace{3mm}}l@{\hspace{3mm}}l}
 \hline\hline
 &&&\multicolumn{8}{c}{${\mathcal W}^{\rm pol}(J_z,J'_z)$}\\
 & & & \multicolumn{4}{c@{\hspace{3mm}}}{linear polarization} &
 \multicolumn{4}{c@{\hspace{3mm}}}{left circular polarization}
 \\
 \cline{4-11}
 $n$ & $J_z$ & $J'_z$ & \hspace*{-3mm}$\beta=0$ &
 $\pi/4$ & $\pi/3$ & $\pi/2$ &
 $0$ & $\pi/4$ & $\pi/3$ & $\pi/2$ \\
 \hline
 1 & \phantom{$-$}1/2 & \phantom{$-$}3/2 &
 0.1250 & 0. & 0.0312 & 0.1250 &
 0. & 0.0312 & 0.0625 & 0.0625 \\
 2 & \phantom{$-$}1/2 & \phantom{$-$}1/2 &
 0. & 0.3750 & 0.2813 & 0. &
 0. & 0.1875 & 0.1406 & 0.\\
 3 & \phantom{$-$}1/2 & $-$1/2 &
 0.3750 & 0. & 0.0938 & 0.3750 &
 0.7500 & 0.0938 & 0. & 0.1875 \\
 4 & \phantom{$-$}1/2 & $-$3/2 &
 0. & 0.1250 & 0.0938 & 0. &
 0. & 0.3643 & 0.4218 & 0.2500 \\
 5 & $-$1/2 & \phantom{$-$}3/2 &
 0. & 0.1250 & 0.0938 & 0. &
 0. & 0.0107 & 0.0469 & 0.2500 \\
 6 & $-$1/2 & \phantom{$-$}1/2 &
 0.3750 & 0. & 0.0938 & 0.3750 &
 0. & 0.0938 & 0.1875 & 0.1875 \\
 7 & $-$1/2 & $-$1/2 &
 0. & 0.3750 & 0.2813 & 0. &
 0. & 0.1875 & 0.1406 & 0. \\
 8 & $-$1/2 & $-$3/2 &
 0.1250 & 0. & 0.0312 & 0.1250 &
 0.2500 & 0.0313 & 0. & 0.0625 \\
 \hline
 \end{tabular}
 \end{table}

 The relative intensities of the Zeeman components are described with the
 factor ${\mathcal W}^{\rm pol}(J_z,J'_z)$ in
 Eqs.~(\ref{fact}),(\ref{W^pol}) and strongly depend on geometry and the
 polarization of the incident electromagnetic radiation.
 To investigate this dependence we parameterize the complex unit
 vector $\hat{\bf \epsilon}={\mathbf E}_0/|{\mathbf E}_0|$
 pointing along the electric
 field amplitude ${\mathbf E}_0$ in the following way.
 We denote by $K$ the lab reference frame with $z$-axis along the external magnetic
 field ${\mathbf B}$, by $K'$ a reference frame
 with $z$-axis along $\hat{\bf k}$,
 and take the cartesian coordinates $(\epsilon'_x,\epsilon'_y,\epsilon'_z)$ of
 ${\bf \epsilon}$ in
 $K'$ to be $(\cos\theta, \sin\theta\,e^{i\varphi},0)$.
 Linear polarization of the incident light is described by
 $\varphi=0$; circular polarization -- by
 $\varphi=\pm\pi/2, \theta=\pi/4$; all other combinations
 correspond to general elliptic polarization. Let
 $(\alpha,\beta,\gamma)$ be the Euler angles of the rotation that
 transforms $K$ into $K'$, and denote by $M(\alpha,\beta,\gamma)$
 the matrix relating the cartesian coordinates $(a_x,a_y,a_z)$ and
 $(a'_x,a'_y,a'_z)$ of an arbitrary vector ${\mathbf a}$ in $K$ and
 $K'$, respectively: $a_i=\sum_j M_{ij}(\alpha,\beta,\gamma)\,a'_j$.
 (To avoid mismatch of $M$ with $M^{-1}$,
 note that, e.g. $M_{xz}=-\sin\beta\,\cos\gamma$.)
 In this way, the absolute values of the components of $\widehat{T}$ in the lab frame $K$,
 appearing in Eq.~(\ref{W^pol})
 are expressed in closed form in terms of the four angles
 $\alpha,\beta,\theta$, and $\varphi$ (the dependence on $\gamma$
 being cancelled). Since the general expressions are rather
 lengthy, we restrict ourselves here to the cases of main interest for
 the experiment. We have:\\
 (a) $\varphi = 0$ for linear polarization
 \begin{align}
  |\widehat{T}^{(2)0 \ }|^2&=\frac{1}{4}\sin^22\beta\cos^2(\alpha-\theta),
  \label{linpol}\\
  |\widehat{T}^{(2)\pm 1}|^2&=\frac{1}{12}\left(1+\sin^2(\alpha-\theta)\cos2\beta+
  \cos^2(\alpha-\theta)\cos4\beta\right),\nonumber\\
  |\widehat{T}^{(2)\pm 2}|^2&=\frac{1}{24}\sin^2\beta \,\left(3
  +\cos2\beta-2\sin^2\beta \cos2(\alpha-\theta)\right);\nonumber
 \end{align}
 (b) $\theta =\pi/4$,\
 $\varphi = \pi/2$  for left circular polarization (l.c.p.)
  \begin{align}
  | \ \widehat{T}^{(2) \ 0 \ }|^2&=\frac{1}{8}\sin^22\beta,
  \label{circpol}\\
  |\widehat{T}^{(2) \ \pm 1}|^2&=\frac{1}{3}
   \left(\begin{matrix}
      \sin^4\beta/2 \\
      \cos^4\beta/2
   \end{matrix}\right)
   (1\pm2\cos\beta)^2,\nonumber\\
  |\widehat{T}^{(2) \ \pm 2}|^2&=\frac{1}{3}
   \left(\begin{matrix}
      \sin^4\beta/2 \\
      \cos^4\beta/2
   \end{matrix}\right)
  \sin^2\beta.\nonumber
 \end{align}
  For right circular polarization (r.c.p.), described by $\theta=\pi/4$,
  $\varphi=-\pi/2$, the values of $|\widehat{T}^{(2)q}|^2$ are
  obtained from the above expressions with the substitution
  $|\widehat{T}^{(2)q}(r.c.p.)|^2=|\widehat{T}^{(2)-q}(l.c.p.)|^2$.

 It is worth stressing that the Eqs.~(\ref{fact})-(\ref{W^pol}) are valid for
 arbitrary angle $\beta$ between the magnetic field and the laser
 propagation direction.

\section{\label{sec:Conclusion}Conclusion}

We derived the E2 transition spectrum of H$_2^+$, including the
first systematic consideration of the transition strength and of
the effects of the laser polarization.
The matrix elements of the electric quadrupole transition moment,
needed for the evaluation of the laser-driven transition rates,
have been calculated in a broad spectral range for a very large
number of transitions, using the most advanced computational
methods. The numerical results agree with the results of Refs.
\cite{Pilon2012} and \cite{Karr2014} wherever comparison is
possible.

The results can be used in planning future experiments and in
interpreting the spectroscopy data.
 The most basic application of the results presented here is to
 estimate the
 laser intensity necessary to achieve a desired transition rate.

 The treatment we have given is applicable both to the situation
 when Doppler broadening is present and absent. When it is present,
 then the individual hyperfine components and Zeeman components may
 not be resolved. Several components will contribute to the
 spectroscopic signal even if the laser radiation is perfectly
 monochromatic. The formula for the strengths of the individual components given
 here allows for producing a model of the Doppler-broadened line profile
 which can be used in fitting the experimental signal.
 As the spectroscopy of MHI will develop into the Doppler-free
 regime (Lamb-Dicke regime), the concept of Rabi frequency will
 become more relevant. This can also easily be computed with the
 expressions given here.

 The presented approach is applicable for any
 relative size of the hyperfine coefficients and
 mixing angles $\phi$. Thus, it can
 also be used for molecules, in which the coupling between
 electron spin and rotation (described by the coefficient $c_e$)
 is the strongest, opposite to the case in H$_2^+$.
 An important example which is drawing substantial attention in
 connection with spectroscopy in ion traps, is N$_2^+$ \cite{willi,
 kajita}. Our treatment here is appropriate
 for the ``fermionic'' isotopologue $^{15}{\rm N}_2^+$ with nitrogen
 nuclear spin $1/2$.

\begin{acknowledgments}
D.B. and P.D. gratefully acknowledge the support of the Bulgarian
National Science Fund under Grant No. FNI 08-17, and of the
Bulgarian Academy of Sciences under Grant DFNP-47. D.B. is also
acknowledging the support of a DAAD grant, ref. no. 91618643.
V.I.K. acknowledges support from the Russian Foundation for Basic
Research under Grant No.~15-02-01906-a.
\end{acknowledgments}

\end{document}


\title{Electric quadrupole transition in $H_2^+$ }

\maketitle

\begin{longtable}{c @{\hspace{10mm}}d @{\hspace{1mm}}d @{\hspace{1mm}}d}
  \caption{Non-relativistic E2 transition energies
  $\Delta E^{\rm NR}=E^{\rm NR}_{v'L'}-E^{\rm NR}_{v'L'}$, in $cm^{-1}$,
  reduced matrix elements $\langle v'L'||Q^{(2)}||vL\rangle/ea_0^2$
  of the electric quadrupole moment of H$_2^+$
  between non-relativistic states with $0\le v\le 10$ and $0\le
  L\le 6$, in atomic units $ea_0^2$ ($a_0$ denoting the Bohr radius),
  and Einstein's coefficient $A_{if}$,
  in $s^{-1}$.}\\
 \label{fullt}
 $v,L\rightarrow v',L'$ &
 \multicolumn{1}{c}{
 $\Delta E^{\rm NR}, cm^{-1}\hspace{0mm}$}  &
 \multicolumn{1}{c}{\hspace*{10mm}
 $\langle\ v'L'\bigr\|{\bf Q}^{(2)}\bigr\|vL\rangle$},\ a.u. &
 \multicolumn{1}{c}{\hspace*{10mm}
 $A_{if},\ s^{-1}$}
 \\
 \hline
 \endfirsthead
 \multicolumn{3}{c}%
 {\tablename\ \thetable\ -- \textit{Continued from previous page}} \\
 \hline
 \noindent
 $v,L\rightarrow v',L'$ &
 \multicolumn{1}{c}{
 $\Delta E^{\rm NR}, cm^{-1}\hspace{0mm}$}  &
 \multicolumn{1}{c}{\hspace*{10mm}
 $\langle\ v'L'\bigr\|{\bf Q}^{(2)}\bigr\|vL\rangle$},\ a.u. &
 \multicolumn{1}{c}{\hspace*{10mm}
 $A_{21},\ s^{-1}$}
 \\
 \hline
 \endhead
 \hline \multicolumn{3}{r}{\textit{Continued on next page}} \\
 \endfoot
 \hline
 \endlastfoot
 0,0$\rightarrow$\phantom{5}0,2 &   174.230 &   1.644960 &   0.973137[-11] \\
 0,0$\rightarrow$\phantom{5}1,2 &  2356.154 &  -0.313846 &   0.160207[-06] \\
 0,0$\rightarrow$\phantom{5}2,2 &  4411.180 &  -0.028919 &   0.312877[-07] \\
 0,0$\rightarrow$\phantom{5}3,2 &  6343.488 &  -0.004652 &   0.497870[-08] \\
 0,0$\rightarrow$\phantom{5}4,2 &  8156.596 &  -0.001048 &   0.888671[-09] \\
 0,0$\rightarrow$\phantom{5}5,2 &  9853.380 &  -0.000298 &   0.185072[-09] \\
 0,0$\rightarrow$\phantom{5}6,2 & 11436.088 &  -0.000100 &   0.441771[-10] \\
 0,1$\rightarrow$\phantom{5}0,3 &   288.850 &   2.217064 &   0.158133[-09] \\
\end{longtable}